\documentclass[12pt]{article}
\usepackage{amssymb,graphicx}

\begin{document}
\title{Affine Particles and Fields}

\author{Djordje \v Sija\v cki\thanks{email: sijacki@phy.bg.ac.yu} \\
Institute of Physics, P.O. Box 57, 11001 Belgrade, Serbia}

\date{}

\maketitle

\begin{abstract}
The covering of the affine symmetry group, a semidirect product of
translations and special linear transformations, in $D \geq 3$ dimensional
spacetime is considered. Infinite dimensional spinorial representations on
states and fields are presented. A Dirac-like affine equation, with infinite
matrices generalizing the $\gamma$ matrices, is constructed.
\end{abstract}

\section{Introduction}

Dmitri Ivanenko, one of the great gravitational physicists of the last
century, and G. Sardanashvily begin their comprehensive review paper on gauge
treatment of gravity [1] with the following statement: 
"At present Einstein's General Relativity (GR) still remains the most
satisfactory theory of classical gravitation for all now observable
gravitational fields. GR successfully passed the test of recent experiments
on the radiolocation of planets and on the laser-location of the Moon, which
have put the end to some other versions of gravitation theory, e.g., the
scalar-tensor theory. At the same time the conventional description of
gravity by Einstein's GR obviously faced a number of serious problems,
and even some corner-stones of gravitation theory still remain disputable up
to our day. This is reflected also in the rather curious uninterrupted flow
of proposals for new designations of this theory".

After more than twenty years, despite of important developments in the subject
matter this statement is still fairly accurate. As for the gauge approach to
gravity, the theory obtained by gauging the Poincar\'e group is in a mature
stage. Here, there are definite results concerning the structure of the
theory, classical sector applications, coupling to tensorial and spinorial
matter etc.  However, there are still notable difficulties in the quantum
sector. A sizable part of the above review paper [1] is devoted to the
$GL(4,R)$ symmetry as "one of the most natural candidates to generalize the
Lorentz gauge gravitation". The $SL(4,R) \subset GL(4,R)$ symmetry plays, in
the Affine group case, the role of the Lorentz symmetry in the Poincar\'e
case. A weak point of the metric-affine [2, 3] and/or the gauge-affine [4, 5]
approaches to the gauge theories of gravity is still the one of the spinorial
affine matter description. In particular, there are no yet candidates for a
Dirac-like wave equation that would successfully describe spinorial affine
particles and fields. This is related to the group theoretical properties of
the quantum affine symmetries in $D \geq 3$, especially to the fact that the
corresponding linear spinorial representations are necessarily infinite
dimensional.

The aim of this paper is to shed some light on the description of the 
spinorial affine matter in $D \geq 3$ dimensional spacetime along the lines
of a recent paper that was concerned primarily with the $D = 3$ case [6].
We study the algebraic structure and the construction of the spinorial
representations of the following physically relevant groups,
$$
\matrix{
T_{D}\wedge\overline{SL}(D,R)&\supset&\overline{SL}(D,R)&\supset&Spin(D) \cr 
                    \cup     &       &         \cup     &       &\cup    \cr
 T_{D}\wedge Spin(1,D-1)     &\supset& Spin(1,D-1)&\supset & Spin(D-1)  .\cr}
$$
Moreover, we consider a construction of a Dirac-like equation for
infinite-component spinorial $\overline{SL}(D,R)$ field. This construction is
carried out by embedding the $sl(D,R)$ algebra as well as the corresponding
$D$-vector $X$, that generalizes Dirac's $\gamma$ matrices, into the
$sl(D+1,R)$ algebra. This $D$-dimensional flat-spacetime Dirac-like equations,
for $D \geq 3$, are of significant importance for a construction of spinorial
fields, "world spinors" [7, 8], in a generic non-Riemannian spacetime of
arbitrary torsion and curvature.

\section{Affine group and algebra}

The general affine group $GA(D,R)$, in $D$-dimensional spacetime, is a
semidirect product of the group $T_{D}$ of translations and the general linear
group $GL(D,R)$, i.e. $GA(D,R) = T_{D} \wedge GL(D,R)$. The nontrivial
dimensionality of the $GA(D,R)$ Hilbert space representations is determined by
its special affine subgroup $SA(D,R) = T_{D} \wedge SL(D,R)$. Owing to the
fact that we are interested in nontrivial spinorial and tensorial
affine-spacetime structure, we confine ourselves, in this paper, to the
$SA(D,R)$ group and for the sake of brevity refer to it as to the "affine"
group. The commutation relations of the $sa(D,R)$ algebra of the $SA(D,R)$
group read 
\begin{eqnarray*}
&&[P_{a},\ P_{b}] = 0,\\
&&[Q_{ab},\ P_{c}] = ig_{ac}P_{b},\\
&&[Q_{ab}, Q_{cd}] = ig_{bc}Q_{ad} - ig_{ad}Q_{cb},
\end{eqnarray*}
the structure constants $g_{ab}$ being either
$\delta_{ab} = (+1, +1, ..., +1)$, $a,b,c,d = 1,2,\dots D$ for the $SO(D)$
subgroup or $\eta_{ab} = (+1, -1, ..., -1)$, $a,b,c,d = 0,1,\dots D-1$ for the
$D$-dimensional Lorentz subgroup $SO(1,D-1)$ of the $SL(D,R)$ group. 

The important $sl(D,R)$ subalgebras are as follows.

(i) $so(1,D-1)$: The $M_{ab} = Q_{[ab]}$, for $g_{ab}=\eta_{ab}$, operators
generate the Lorentz-like subgroup $SO(1,D-1)$ with $J_{ij} = M_{ij}$ (angular
momentum) and $K_{i} = M_{0i}$ (the boosts) $i,j = 1,2,\dots ,D-1$.  

(ii) $so(D)$: The $J_{ab} = Q_{[ab]}$, for $g_{ab}=\delta_{ab}$, $J_{ij}$ and
$N_{i} = Q_{\{0i\}}$ operators generate the maximal compact subgroup $SO(D)$.

(iii) $sl(D-1)$: The $J_{ij}$ and $T_{ij} = Q_{\{ij\}}$ operators
generate the subgroup $SL(D-1,R)$ - the "little group of the
massive particle states".

The $SL(D,R)$ commutation relations, in terms of the metric preserving
antisymmetric operators $M_{ab} = Q_{[ab]}$ and the remaining traceless
symmetric operators $T_{ab} = Q_{(ab)}$ that generate the (non-trivial)
$D$-volume preserving transformations, are given as follows
\begin{eqnarray*}
&&[M_{ab},\ M_{cd}] = -i\eta_{ac}M_{bd} +i\eta_{ad}M_{bc}
+i\eta_{bc}M_{ad} -i\eta_{bd}M_{ac}, \\
&&[ M_{ab},\ T_{cd}] = -i\eta_{ac}T_{bd} -i\eta_{ad}T_{bc}
+i\eta_{bc}T_{ad} +i\eta_{bd}T_{ac},\\
&&[ T_{ab},\ T_{cd}] = +i\eta_{ac}M_{bd} +i\eta_{ad}M_{bc}
+i\eta_{bc}M_{ad} +i\eta_{bd}M_{ac} .
\end{eqnarray*}

The quantum mechanical symmetry group $G_{qm}$ is given as the $U(1)$ minimal
extensions of the corresponding classical symmetry group $G_{cl}$,     
$$
1 \rightarrow U(1) \rightarrow G_{qm} \rightarrow G_{cl} \rightarrow 1 \ . 
$$
In practice, finding $G_{qm}$ is accounted for by taking the universal
covering group of the $G_{cl}$ group (topology changes), and by solving the
algebra commutation relations for possible central charges (algebra
deformation). There are no nontrivial central charges of the $sa(D,R)$ and
$sl(D,R)$ algebras, and the remaining important question for quantum
applications is the one of the affine symmetry covering group. The
translational part of the $SA(D,R)$ group is contractible to a point and thus
irrelevant for the covering question. The $SL(D,R)$ subgroup is, according to
the Iwasawa decomposition, given by $SL(D,R)$ $=$ $SO(D,R)\times A\times N$,
where $A$ is a subgroup of Abelian transformations (e.g. diagonal matrices)
and $N$ is a nilpotent subgroup (e.g., upper triangular matrices). Both $A$
and $N$ subgroups are contractible to point. Therefore, the covering features
are determined by the topological properties of the maximal compact subgroup of
the group in question. In our case, that is the $SO(D,R)$ group, i.e. more
precisely its central subgroup. 

The universal covering group of the $SO(D)$, $D\geq 3$ group is its double
covering group isomorphic to $Spin(D)$. In other words $SO(D)$ $\simeq$
$Spin(D)/Z_{2}$ (for a detail account of $Spin(D)$ and $Pin(D)$ groups
cf. [9]). The special affine and the special linear groups have
double-coverings as their universal coverings as summarized by the following
diagram of exact sequences
$$
\matrix{
   & &  &  &  1     &  & 1     &  &  \cr
   &  &  &  & Z_{2} &  & Z_{2} &  &  \cr
   &  &  &  & \downarrow &  & \downarrow &  &  \cr
 1 & \rightarrow & T_{D} & \rightarrow & \overline{SA}(D,R) &
\rightarrow & \overline{SL}(D,R) & \rightarrow & 1 \cr
   &  &  &  & \downarrow &  & \downarrow &  &  \cr
 1 & \rightarrow & T_{D} & \rightarrow & SA(D,R) &
\rightarrow & SL(D,R) & \rightarrow & 1 \cr
   &  &  &  & \downarrow &  & \downarrow &  &  \cr
   &  &  &  & \downarrow &  & \downarrow &  &  \cr
   &  &  &  & 1 &  & 1 &  &  \cr}
$$ 

In the physically most interesting case, $D=4$, there is a
homomorphism between $SO(3)\times SO(3)$ and $SO(4)$. Since $SO(3)
\simeq SU(2)/Z_{2}$, where $Z_{2}$ is the two-element center $\{
1,-1\}$, one has $SO(4) \simeq [SU(2)\times SU(2)]/Z_{2}^{d}$, where
$Z_{2}^{d}$ is the diagonal discrete group whose representations are
given by $\{ 1,(-1)^{2j_{1}}=(-1)^{2j_{2}} \}$ with $j_{1}$ and
$j_{2}$ being the Casimir labels of the two $SU(2)$ representations. The
full $Z_{2}\times Z_{2}$ group, given by the representations $\{
1,(-1)^{2j_{1}} \} \otimes \{ 1,(-1)^{2j_{2}} \}$, is the center of
$Spin(4)=SU(2)\times SU(2)$, which is thus the
quadruple-covering of $SO(3)\times SO(3)$ and a double-covering of
$SO(4)$. The groups $SO(3)\times SO(3)$, $SO(4)$ and
$Spin(4) \simeq SU(2)\times SU(2)$ are thus the maximal compact
subgroups of $SO(3,3)$, $SL(4,R)$ and $\overline{SL}(4,R)$
respectively. One can sum up these results by the following exact
sequences
$$
\matrix{
   & &  &  &     1 &  &            1 &  &  \cr
   &  &  &  & \downarrow &  & \downarrow &  &  \cr
 1 & \rightarrow & Z_{2}^{d} & \rightarrow & Z_{2}\times Z_{2} &
\rightarrow & Z_{2} & \rightarrow & 1 \cr
   &  &  &  & \downarrow &  & \downarrow &  &  \cr
 1 & \rightarrow & Z_{2}^{d} & \rightarrow & \overline{SL}(4,R) &
\rightarrow & SL(4,R) & \rightarrow & 1 \cr
   &  &  &  & \downarrow &  & \downarrow &  &  \cr
   &  &  &  & SO(3,3) &  & SO(3,3) &  &  \cr
   &  &  &  & \downarrow &  & \downarrow &  &  \cr
   &  &  &  & 1 &  & 1 &  &  \cr}
$$

The universal covering group $\overline{G}$ of a given group $G$ is a group
with the same Lie algebra and with a simply-connected group manifold. A finite
dimensional covering, $\overline{SL}(D,R)$, exists provided one can embed
$SL(D,R)$ into a group of finite complex matrices that contain $Spin(D)$ as
subgroup. A scan of the Cartan classical algebras points to the $SL(D,C)$
groups as a natural candidate for the $SL(D,R)$ groups covering. However,
there is no match of the defining dimensionalities of the $SL(D,R)$ and
$Spin(D)$ groups for $D\geq 3$, 
$$
dim(SL(D,C)) = D \quad < \quad 2^{\left[\frac{D-1}{2}\right] } = 
dim(Spin(D)) ,
$$
except for $D=8$. In the $D=8$ case, one finds that the orthogonal subgroup of
the $SL(8,R)$ and $SL(8,C)$ groups is $SO(8,R)$ and not $Spin(8)$. For a
detailed account of the $D=4$ case cf. [10]. Thus, we conclude that
there are no finite-dimensional covering groups of the $SL(D,R)$ groups for
any $D \geq 3$. An explicit construction of all spinorial, unitary and
nonunitary multiplicity-free [11] and unitary non-multiplicity-free
[12], $SL(3,R)$ representations shows that they are all defined in
infinite-dimensional spaces. 

The universal (double) covering groups of the $\overline{SL}(D,R)$ and
$\overline{SA}(D,R)$, $D \geq 3$ groups are groups of infinite complex
matrices. All their spinorial representations are infinite-dimensional and
when reduced w.r.t. $Spin(D)$ subgroups contain representations of unbounded
spin values.

\section{Representations on states}

The $\overline{SA}(D,R)$ Hilbert space representations are, owing to the
semidirect product group structure, induced as in the Poincar\'e case
from the corresponding little group (stability subgroup)
representations. The correct quantum mechanical interpretation requires these
representations to be unitary.

The steps in the construction of the unitary irreducible $\overline{SA}(D,R)$
Hilbert space representations are, in the quantum physics terminology, as
follows: (i) determine the vectors characterized by the maximal set of good
quantum numbers of the Abelian subgroup $T_{D}$ generators, (ii) 
determine the little group as a subgroup of the $SL(D,R)$ transformations that
leaves these vectors invariant, and (iii) induce the unitary irreducible
$\overline{SA}(D,R)$ representations from the corresponding $T_{D}$ and little
group representations. In contradistinction to the Poincar\'e case, the little
groups that describe affine particles are more complex in structure due to the
fact that a orthogonal type of group is enlarged here to the linear one.

The little group of the $\overline{SA}(D,R)$ Hilbert-space particle states is
of the form $T^{\sim}_{D-1}\wedge\overline{SL}(D-1,R)$, where the Abelian
invariant subgroup $T^{\sim}_{D-1}$ of the little group is generated by
$Q_{1j}$, $j=2,3,\dots ,D$. Owing to the fact that the little group is itself
given as a semidirect product, we have the following possibilities:

(i) The whole little group is represented trivially corresponding to a scalar
state.

(ii) The $T^{\sim}_{D-1}$ subgroup is represented trivially,
$D(T^{\sim}_{D-1}) \to 1$, i.e. $D(Q_{1j})$ $\to$ $0$, the remaining
little group is $\overline{SL}(D-1,R)$, and the corresponding "affine
particle" is described by the unitary irreducible $\overline{SL}(D-1,R)$
representations. These representations are infinite dimensional, even in the
tensorial case, due to noncompactness of the $SL(D,R)$ group. 

(iii) The whole little group $T^{\sim}_{D-1}\wedge\overline{SL}(D-1,R)$ is
represented nontrivially. The corresponding "affine particles" are described
by $D-1$ real additive quantum numbers provided by $Q_{1k}$, $k=2,3,\dots ,D$,
and the representations of a next step little group
$T^{\sim}_{D-2}\wedge\overline{SL}(D-2,R)$ that is a subgroup of
$\overline{SL}(D-1,R)$. The $T^{\sim}_{D-2}$ subgroup is generated by
$Q_{2k}$, $k=3,4,\dots ,D$. Here, we have again the above branching situation,
either we represent $T^{\sim}_{D-2}$ trivially and have an effective
$\overline{SL}(D-2,R)$ little group, or we represent $T^{\sim}_{D-2}$
nontrivially and arrive at the next step little group
$T^{\sim}_{D-3}\wedge\overline{SL}(D-3,R)$ $\subset$ $\overline{SL}(D-2,R)$,
and so on.
$$
\matrix{
\overline{SL}(D,R)   & & \cr
\cup      & & \cr
T^{\sim}_{D-1}\wedge\overline{SL}(D-1,R) & \supset & \overline{SL}(D-1,R) \cr
\cup      & & \cr
T^{\sim}_{D-2}\wedge\overline{SL}(D-2,R) & \supset & \overline{SL}(D-2,R) \cr
\cup      & & \cr
\cdots    & & \cr
\cup      & & \cr
T^{\sim}_{3}\wedge\overline{SL}(3,R) & \supset & \overline{SL}(3,R) \cr
\cup      & & \cr
T^{\sim}_{2}\wedge\overline{SL}(2,R) & \supset & \overline{SL}(2,R) \cr}
$$

Let $(a,\bar A) \in \overline{SA}(D,R)$, $a\in T_{D}, \bar A\in
\overline{SL}(D,R)$ be the group elements. The group composition law is
$$
(a_{1},\bar A_{1})(a_{2},\bar A_{2})=(a_{1}+A_{1}a_{2},\bar A_{1}\bar A_{2}),
$$
where $A\in SL(D,R)$ corresponds to $\bar A\in\overline{SL}(D,R)$ through
$\overline{SL}(4,R)/Z_{2} \to SL(D,R)$ .

In the case when $T^{\sim}_{D-1}$ is represented trivially, the
$\overline{SA}(D,R)$ representations on states are given by the following
expression
$$
D(a,\bar A) f^{[j]}(p,[m])
= e^{ia\cdot (Ap)} \sum\limits_{[m^{\prime}]} D^{[j]}_{[m^{\prime}][m]}
(L_{Ap}^{-1}\bar AL_{p})f^{[j]}(Ap,[m^{\prime}]),
$$
where $[j]$ are the $\overline{SL}(D,R)$ quantum numbers, and $L_{p}$
represents the action of an element $C$ defined by $\bar A=C\bar H,
H\in SL(D-1,R)$ on the state $p_{(0)} =(p_{0},0,0,0)$, i.e., $p=L_{p} p_{(0)}
=C p_{(0)} $. 
The $\overline{SL}(D-1,R)$ subgroup of the $\overline{SA}(D,R)$ group is
represented linearly, while the elements of the $\overline{SL}(D,R) / 
\overline{SL}(D-1,R)$ factor group are primarily realized nonlinearly over
$SL(D-1,R)$ and then represented linearly. Once again, the remaining
$\overline{SL}(D-1,R)$ little group is noncompact, and in the quantum case one
has to make use of its unitary irreducible representations (both spinorial and
tensorial) that are necessarily infinite dimensional.

\section{Representations on fields}

The representations of the Poincar\'e group on fields are given by
the following well known expressions
\begin{eqnarray*}
&&(D(a,\bar\Lambda )\Phi_m)(x) = (D(\bar\Lambda ))^n_m \Phi_n
(\Lambda^{-1}(x-a)) \\ 
&&(a, \bar\Lambda ) \in T_{D}\wedge Spin(1,D-1) \nonumber ,
\end{eqnarray*}
where $m,n$ enumerate a basis of the representation space of the field
components. 

In the standard applications to gravity and/or particle physics, one makes use
of the finite-component representations of the Lorentz group on fields. This
is in agreement with experiment. For instance, boosted particles do not get
spin excited. The fact that finite-dimensional representations,
$D(\bar\Lambda{})$,  
of the Lorentz subgroup are, due to its noncompactnes, nonunitary is of no
physical relevance. In fact, only the field components corresponding to the
modes described by the unitary representation of the little group are allowed
to propagate by means of field equations. In other words, unitarity is
imposed in the Hilbert space of the representations on states only, while the
field equations provide for a full Lorentz covariance, and restrict the
field components in such a way that the physical degrees of freedom are as
given by the corresponding particle states.

Representations of the affine group $\overline{SA}(D,R)$ on fields are given
by the same expression with the Lorentz group being replaced by the
$\overline{SL}(D,R)$ group. There are two physical requirements that have to
be satisfied in the affine case: (i) representations of the Lorentz subgroup
$Spin(1,D)$ have to be finite-dimensional and thus nonunitary and (ii)
representations of the affine-particle little group $\overline{SL}(D-1,R)$
have to be unitary and thus (due to little group's noncompactnes)
infinite-dimensional.  

The correct unitarity properties of the affine fields can be achieved by
making use of the unitary (irreducible) representations and the so called
"deunitarizing" automorphism of the $\overline{SL}(D,R)$ group. The
$\overline{SL}(D,R)$ commutation relations are invariant under the
``deunitarizing'' automorphism [10], 
\begin{eqnarray*}
&&{\cal A}: \ \overline{SL}(D,R) \rightarrow \overline{SL}(D,R) \\
&&J^{\cal A}_{ij} = J_{ij} ,\quad K^{\cal A}_{j} = iN_{j} ,\quad
N^{\cal A}_{j} = iK_{j} , \\
&&T^{\cal A}_{ij} = T_{ij} ,\quad
T^{\cal A}_{00} = T_{00} ,\quad i,j=1,2,\dots ,D-1, \\
\end{eqnarray*}
so that $(J_{ij},\ iK_{i})$ generate the new compact $Spin(D)^{\cal A}$ and
$(J_{ij},\ iN_{i})$ generate $Spin(1,D-1)^{\cal A}$.

For the (spinorial) particle states, we use the basis vectors of the
unitary irreducible representations of $\overline{SL}(D,R)^{\cal A}$, so that
the compact subgroup finite multiplets correspond to $Spin(D)^{\cal A}$,
generated by $\{ J_{ij},\ iK_{i}\}$, while $Spin(1,D-1)^{\cal A}$, generated
by $\{ J_{ij},\ iN_{j}\}$, is represented by unitary infinite-dimensional
representations. We now perform the inverse transformation and return to
$\overline{SL}(D,R)$ for our physical identification. $\overline{SL}(D,R)$ is
represented non-unitarily, the compact $Spin(D)$ is represented by non-unitary
infinite representations while the Lorentz group is represented by non-unitary
finite representations. These finite-dimensional non-unitary Lorentz group
representations are precisely those that ensure a correct particle
interpretation. Note that $\overline{SL}(D-1,R)$, the stability subgroup of
$\overline{SA}(D,R)$, is represented unitarily.

We now face the problem of constructing the (unitary) infinite-dimensional
spinorial and tensorial representations of the $\overline{SL}(D,R)$ group.
The $\overline{SL}(D,R)$ group can be contracted (a la Wigner-In\"on\"u)
w.r.t. its $Spin(D)$ subgroup to yield the semidirect-product group $T'\wedge
Spin(D)$. $T'$ is an $\frac{1}{2}(D+2)(D-1)$ parameter Abelian group generated
by operators $U_{ab}$ $=$ $\lim_{\varepsilon\to 0} (\varepsilon T_{ab})$,
which form a $Spin(D)$ second rank symmetric operator obeying the following
commutation relations, 
\begin{eqnarray*}
&&[J_{ab},\ J_{cd}] = -i\eta_{ac}J_{bd} +i\eta_{ad}J_{bc}
+i\eta_{bc}J_{ad} -i\eta_{bd}j_{ac}, \\
&&[ J_{ab},\ U_{cd}] = -i\eta_{ac}U_{bd} -i\eta_{ad}U_{bc}
+i\eta_{bc}U_{ad} +i\eta_{bd}U_{ac},\\
&&[ U_{ab},\ U_{cd}] = 0 .
\end{eqnarray*}

An efficient way of constructing explicitly the $\overline{SL}(D,R)$
infinite-dimen\-si\-onal representations consists in making use of the
so called "decontraction" formula, which is an inverse of the
Wigner-In\"on\"u contraction. According to the decontraction formula,
the following operators 
$$
T_{ab} = p U_{ab} + \frac{i}{2\sqrt{U\cdot U}} 
\left[ C_2(Spin(D)),\ U_{ab} \right],
$$
together with $J_{ab}$ form the $\overline{SL}(D,R)$ algebra. The parameter
$p$ is an arbitrary complex number, $p\in C$, and $C_2(Spin(D))$ is the
$Spin(D)$ second-rank Casimir operator.
 
For the representation Hilbert space we take the homogeneous space of
$L^2$ functions of the maximal compact subgroup $Spin(D)$ parameters. The
$Spin(D)$ representation labels are given either by the Dynkin labels
$({\lambda}_1, {\lambda}_2,\dots , {\lambda}_r)$ or by the highest weight
vector which we denote by $\{ j\} = \{ j_1, j_2, \dots , j_r \}$, $r=
\left[\frac{D}{2}\right]$. 
The $\overline{SL}(D,R)$ commutation relations are invariant w.r.t. an
automorphism defined by:
$$
s(J) = +J, \quad\quad s(T) = -T .
$$ 
This allows us to associate an '$s$-parity' to each $Spin(D)$ representation
of an $\overline{SL}(D,R)$ representation. In terms of Dynkin labels we find
\begin{eqnarray*}
s(D_2) &=& (-)^{{1 \over 2}({\lambda}_1 + {\lambda}_2 - \epsilon )},
\\ s(D_{n\ge 3}) &=& (-)^{ {\lambda}_1 + {\lambda}_2 + \dots +
{\lambda}_{n-2} + {1\over 2}({\lambda}_n - {\lambda}_{n-1} - \epsilon
)} \\ s(B_1) &=& (-)^{ {1\over 2}({\lambda}_1 - \epsilon )} \\
s(B_{n\ge 2}) &=& (-)^{ {\lambda}_1 + {\lambda}_2 + \dots +
{\lambda}_{n-1} + {1\over 2}({\lambda}_n - \epsilon )} \\
\end{eqnarray*}
where $\epsilon = 0$ and $\epsilon =1$ for $\lambda$ even and odd,
respectively. 

The $s$-parity of the $\frac{1}{2}(D+2)(D-1)$-dimension representation
$(20\dots 0)$ $=$ $\Box\!\Box$ of $Spin(D)$ is $s(\Box\!\Box )= +1$. A basis of
an $Spin(D)$ irreducible representation is provided by the Gel'fand-Zetlin
pattern characterized by the maximal weight vectors of the subgroup chain
$Spin(D)$ $\supset$ $Spin(D-1)$ $\supset$ $\cdots$ $\supset$ $Spin(2)$. We
write the basic vectors as $\left| { \{ j \}   }\atop { \{m \} }\right>$,
where $\{ m \}$ corresponds to $Spin(D-1)$ $\supset$ $Spin(D-2)$ $\supset$
$\cdots$ $\supset$ $Spin(2)$ subgroup chain weight vectors.

The Abelian group generators $\{ U\}$ $=$ $\{ U_{[\mu ]}^{[\Box\!\Box ]}\}$
can be, in the case of multiplicity free representations, written in
terms of the $Spin(D)$-Wigner functions as follows $U_{[\mu
]}^{[\Box\!\Box ]}$ $=$ $D_{[0][\mu ]}^{[\Box\!\Box ]}(\phi )$. It is
now rather straightforward to determine the noncompact operators matrix
elements, which are given by the following expression [13]:
\begin{eqnarray*}
\left< \begin{array}{c} \{ j'\} \\ \{ m'\} \end{array} \right|  T_{\{
\mu \}}^{\{ \Box\!\Box \}}  \left| \begin{array}{c} \{ j\} \\ \{ m\}
\end{array} \right>  &=& \left( \begin{array}{ccc} \{ j'\} & \{
\Box\!\Box \} & \{ j\} \\ \{ m'\} & \{ \mu \} & \{ m\} \end{array}
\right)  \left< \{ j'\} \right|| T^{\{ \Box\!\Box \} } \left|| \{ j \}
\right>,   \\ 
\left< \{ j'\} \right|| T^{\{ \Box\!\Box \} } \left|| \{
j\} \right> &=& \sqrt{dim\{ j'\} dim\{ j\} }  \left\{ p + \frac{1}{2}
( C_2 (\{ j'\}) - C_2 (\{ j\} ) ) \right\}  \\
&&\times \left(\begin{array}{ccc} \{ j'\} & \{ \Box\!\Box \} & \{ j\} \\ 
\{ 0\} & \{  0\} &  \{ 0\} \end{array}   \right). \\
\end{eqnarray*}
$\pmatrix{\cdot&\cdot&\cdot\cr \cdot&\cdot&\cdot\cr}$ is the
appropriate $"3j"$ symbol for the $Spin(D)$ group. The (unitary)
infinite-dimensional representations of the $\overline{SL}(D,R)$ algebra are
given by these expressions of the non-compact generators together with the
well known expressions for the maximal compact $Spin(D)$ algebra
representations. Finally, we apply the deunitarizing automorphism $\cal A$ for
a correct physical interpretation.

In the case of the multiplicity free $\overline{SL}(D,R)$ representations, each
$Spin(D)$ sub-representation appears at most once and has the
same $s$-parity. This feature is especially useful for the task of reducing
infinite-dimensional spinorial and tensorial representations of the
$\overline{SL}(D,R)$ group to the corresponding $\overline{SL}(D-1,R)$
sub-representations.

\section{Spinorial wave equations}

Let us consider the question of constructing a Dirac-like equation for an
infinite-component spinorial affine field $\Psi (x)$, 
\begin{eqnarray*}
&&(iX^{a}\partial_{a} -M) \Psi (x) = 0, \\
&&\Psi (x) \sim D_{spin}(\overline{SL}(D,R)) .
\end{eqnarray*}
The $X^{a}$, $a=0,1,\dots ,D-1$ vector operator, acting in the space
of the $\Psi$ field components, is an appropriate generalization of
the Dirac $\gamma$ matrices to the affine case. The
$\overline{SL}(D,R)$  affine covariance requires that the following
commutation relations are satisfied 
\begin{eqnarray*}
&&[M_{ab}, X_{c}] = i \eta_{bc} X_{a} - i \eta_{ac} X_{b} \\ 
&&[T_{ab}, X_{c}] = i \eta_{bc} X_{a} + i \eta_{ac} X_{b} .
\end{eqnarray*}
The first relation ensures Lorentz covariance, and is a easy one to
fulfill. The second relation, required by the full affine covariance, turns
out to be rather difficult to accomplish (cf. [14]).

One can obtain the matrix elements of the generalized Dirac matrices
$X_{a}$ by solving the above commutation relations for $X_{a}$ in the Hilbert
space of a suitable spinorial $\overline{SL}(D,R)$
representation. Alternatively, one can embed both the $\overline{SL}(D,R)$
algebra and $X_{a}$ into the $\overline{SL}(D+1,R)$ algebra, and make use of
representations of the embedding algebra to solve for $X_{a}$. Let us
denote the generators of $\overline{SL}(D+1,R)$ by $Q^{(D+1)}_{AB}$, $A,B =
0,...,D$. Now, there are two natural $D$-vector candidates for $X_{a}$ in
$\overline{SL}(D+1,R)$, i.e. $A_{a}$, and $B_{a}$ defined by 
\begin{eqnarray*}
A_{a} = Q^{(D+1)}_{aD}, \quad B_{a} = Q^{(D+1)}_{Da}, \quad a=0,1,\dots ,D-1 .
\end{eqnarray*}
The operators $A_{a}$ and $B_a$, obtained in this way, fulfill the required
$SL(D,R)$ $D$-vector commutation relations by construction. It is
interesting to note that the operator $G_{a} = \frac{1}{2}(A_{a}-B_{a})$
satisfies
\begin{eqnarray*}
[G_{a}, G_{b}] = -i M_{ab} ,
\end{eqnarray*}
thereby generalizing the corresponding property of Dirac's
$\gamma$-matrices. Since $X_{a}$, $M_{ab}$ and $T_{ab}$ form a closed algebra,
the $X_{a}$ operator connects only those $\overline{SL}(D,R)$
representation states that are contained in the $\overline{SL}(D+1,R)$
representation Hilbert space. By reducing a spinorial $\overline{SL}(D+1,R)$
representation to the $\overline{SL}(D,R)$ sub representations, we obtain a
set of these representations that is closed w.r.t. an $X_{a}$
action. Moreover, an explicit form of the $X_{a}$ matrix elements is
provided by the $\overline{SL}(D+1,R)$ representation expressions.

There are quite a number of substantial changes when going from the Poincar\'e
to the affine symmetry: spinorial representations are infinite dimensional,
unitarity requirements are different, tensor algebra relevant for the wave
equation questions is more restrictive etc. In order to have an impression
about the general structure of the $X_{a}$ matrix, let us consider a toy model,
where we make use of the finite-dimensional tensorial $SL(D,R)$
representations. As an example, let as start with the following
$\frac{1}{6}(D+1)(D+2)(D+3)$-dimensional tensorial irreducible representation
of $SL(D+1,R)$ that reduces to four $SL(D,R)$ representations as follows,
\begin{eqnarray*}
SL(D+1,R) 
&\supset& SL(D,R) \nonumber\\ 
\stackrel{\varphi_{ABC}}{ 
\begin{picture}(26,8)
\put(1,-1){\framebox(24,8)} 
\put(9,-1){\line(0,1){8}}
\put(17,-1){\line(0,1){8}}
\end{picture}}
&\supset& 
\stackrel{\varphi_{abc}}{
\begin{picture}(26,8)
\put(1,-1){\framebox(24,8)} 
\put(9,-1){\line(0,1){8}}
\put(17,-1){\line(0,1){8}}
\end{picture} }
\oplus 
\stackrel{\varphi_{ab}}{
\begin{picture}(26,8)
\put(1,-1){\framebox(24,8)}
\put(9,-1){\line(0,1){8}} 
\put(17,-1){\line(0,1){8}}   
\put(1,0){$\times$}
\end{picture}} 
\oplus 
\stackrel{\varphi_{a}}{
\begin{picture}(26,8)
\put(1,-1){\framebox(24,8)}
\put(9,-1){\line(0,1){8}}
\put(17,-1){\line(0,1){8}}
\put(1,0){$\times$}
\put(9,0){$\times$} 
\end{picture}} 
\oplus 
\stackrel{\varphi}{
\begin{picture}(26,8)
\put(1,-1){\framebox(26,8)}
\put(9,-1){\line(0,1){8}}
\put(17,-1){\line(0,1){8}}
\put(1,0){$\times$}
\put(9,0){$\times$}
\put(17,0){$\times$}
\end{picture}} \ ,
\end{eqnarray*}
where "box" is the Young tableau for an irreducible vector representation
of $SL(D,R)$. The effect of the action of the $SL(D,R)$
vector $X_{a}$ on the fields $\varphi$, $\varphi_{a}$ and $\varphi_{ab}$ and
$\varphi_{abc}$ is
\begin{eqnarray*}
&&\stackrel{X_a}{
\begin{picture}(10,8)
\put(1,-1){\framebox(8,8)}
\end{picture}}
\otimes 
\stackrel{\varphi}{
\begin{picture}(26,8)
\put(1,-1){\framebox(26,8)}
\put(9,-1){\line(0,1){8}}
\put(17,-1){\line(0,1){8}}
\put(1,0){$\times$}
\put(9,0){$\times$}
\put(17,0){$\times$}
\end{picture}}
\ \mapsto\  
\stackrel{\varphi_a}{
\begin{picture}(26,8)
\put(1,-1){\framebox(24,8)}
\put(9,-1){\line(0,1){8}}
\put(17,-1){\line(0,1){8}}
\put(1,0){$\times$}
\put(9,0){$\times$} 
\end{picture}}, 
\quad\quad
\stackrel{X_a}{
\begin{picture}(10,8)
\put(1,-1){\framebox(8,8)}
\end{picture}}
\otimes 
\stackrel{\varphi_a}{
\begin{picture}(26,8)
\put(1,-1){\framebox(24,8)}
\put(9,-1){\line(0,1){8}}
\put(17,-1){\line(0,1){8}}
\put(1,0){$\times$}
\put(9,0){$\times$} 
\end{picture} }
\ \mapsto\  
\stackrel{\varphi_{ab}}{
\begin{picture}(26,8)
\put(1,-1){\framebox(24,8)}
\put(9,-1){\line(0,1){8}} 
\put(17,-1){\line(0,1){8}}   
\put(1,0){$\times$}
\end{picture}} , \\
&&
\stackrel{X_a}{
\begin{picture}(10,8)
\put(1,-1){\framebox(8,8)}
\end{picture}}
\otimes \stackrel{\varphi_{ab}}{
\begin{picture}(26,8)
\put(1,-1){\framebox(24,8)}
\put(9,-1){\line(0,1){8}} 
\put(17,-1){\line(0,1){8}}   
\put(1,0){$\times$}
\end{picture}} 
\ \mapsto\ 
\stackrel{\varphi_{abc}}{
\begin{picture}(26,8)
\put(1,-1){\framebox(24,8)} 
\put(9,-1){\line(0,1){8}}
\put(17,-1){\line(0,1){8}}
\end{picture} } ,
\quad\quad
\stackrel{X_a}{
\begin{picture}(10,8)
\put(1,-1){\framebox(8,8)}
\end{picture}}
\otimes \stackrel{\varphi_{abc}}{
\begin{picture}(26,8)
\put(1,-1){\framebox(24,8)}
\put(9,-1){\line(0,1){8}} 
\put(17,-1){\line(0,1){8}}
\end{picture}} 
\ \mapsto\  0 \,.
\end{eqnarray*}

Other possible Young tableaux do not appear due to the tensor algebra of the
chosen $\overline{SL}(D+1,R)$ representation. Gathering these fields in a
vector $\Phi = (\varphi, \varphi_{a}, \varphi_{ab},
\varphi_{abc})^{\rm T}$, we can read off the matrix structure of $X_{a}$.

It is interesting to observe here that $X_{a}$ has zero matrices on
the block-diagonal which implies that the mass operator $M$ in an
affine invariant equation vanishes. Consider now an action of the
$X_{a}$ vector operator on an arbitrary irreducible 
representation $D(g)$ of $SL(D,R)$ labeled by
$[\nu_{1},\nu_{2},\dots \nu_{D-1}]$, $\nu_{i}$ being
the number of boxes in the $i$-th raw,  
\begin{eqnarray*}
&&[\nu_{1}, \nu_{2},\dots ,\nu_{D-1}] \otimes [1,0,\dots ,0] \\
&&= [\nu_{1}+1, \nu_{2},\dots ,\nu_{D-1}] \oplus
  [\nu_{1}, \nu_{2}+1,\dots ,\nu_{D-1}] \oplus\dots \\
&&\oplus [\nu_{1}, \nu_{2},\dots ,\nu_{D-1}+1]
  \oplus [\nu_{1}-1, \nu_{2}-1,\dots ,\nu_{D-1}-1] ,
\end{eqnarray*}
where one counts, on the right hand side, the allowed representations only.
None of the resulting representations is isomorphic to the starting
representation $D(g)$. This implies zero matrices on the
block-diagonal of $X_{a}$, in the Hilbert space of an arbitrary $SL(D,R)$ 
irreducible representation. Let the representation space of an
arbitrary reducible representation be spanned by
$\Phi = (\varphi_{1}, \varphi_{2},\dots )^{\rm T}$ with $\varphi_{i}$
irreducible. 
Now we consider the Dirac-type equation in the rest frame
$p_{(0)} = (p_{0},0,\dots ,0)$, restricted to the subspaces spanned by
$\varphi_{i}$, $(i=1,2,\dots )$, 
$$
p_{0} <\varphi_{i}| X^{0} |\varphi_{j}>  -  
 <\varphi_{i}| M |\varphi_{i} > \delta_{ij} = 0 ,
$$
where we assumed the operator $M$ to be diagonal. It follows that the
$M$ operator vanishes  since $<\varphi_{i}|X^{0}|\varphi_{i}> =
0$. 

Let us now turn to the proper spinorial case of infinite-dimensional spinorial
representations of the $\overline{SL}(D,R)$ group. We embed the
$\overline{SL}(D,R)$ algebra, as well as the Dirac-like wave equation
$D$-vector operator $X^{a}$, $a = 0, 1,\dots ,D-1$, into the
$\overline{SL}(D+1,R)$ algebra, and thus satisfy the $[Q_{ab}, X_{c}]$
commutation relations by construction. Moreover, this embedding puts
a constraint on the set of $\overline{SL}(D,R)$ spinorial representations
which define a Hilbert space of field components that is invariant
w.r.t. $X_{a}$ action. 

An explicit construction consists of: (i) a construction of the unitary
spinorial $\overline{SL}(D+1,R)$ representations, (ii) an application of the
deunitarizing $\cal A$ operator, (iii) an identification of the relevant
physical operators in the $\overline{SL}(D+1,R)$ algebra, (iv) a reduction of
the chosen $\overline{SL}(D+1,R)$ spinorial representation down to the
corresponding $\overline{SL}(D,R)$ sub-representations, and (v) an evaluation
of the $X^{a}$ matrix elements in the $\overline{SL}(D,R)$ representation
basis starting with the $\overline{SL}(D+1,R)$ representation matrix elements.

We repeat first the construction procedure, developed for the
$\overline{SL}(D,R)$ representations on fields, but this time in the
$\overline{SL}(D+1,R)$ case. In this way, we arrive at the following
expressions for the compact, $J^{(D+1)}_{AB}$ $=$ $Q^{(D+1)}_{[AB]}$, and the
noncompact, $T^{(D+1)}_{AB}$ $=$ $Q^{(D+1)}_{\{ AB\}}$, generators of the
$\overline{SL}(D+1,R)$ group in the $Spin(D+1)$ representations basis, 
\begin{eqnarray*}
&&\left< \begin{array}{c} \{ j'\} \\ \{ m'\} \end{array} \right|  
J_{\{\mu \}}^{(D+1)\{ \Box \}}  
\left| \begin{array}{c} \{ j\} \\ \{ m\} \end{array} \right>  
= \sqrt{ dim\{ j\}}
\left( \begin{array}{ccc} \{ j' \} & \{\Box \} & \{ j \} \\ 
\{ m'\} & \{ \mu \} & \{ m\} \end{array} \right) \delta_{\{ j'\} \{ j\}}, 
\\
&&\left< \begin{array}{c} \{ j'\} \\ \{ m'\} \end{array} \right|  T_{\{
\mu \}}^{(D+1)\{ \Box\!\Box \}}  \left| \begin{array}{c} \{ j\} \\ \{ m\}
\end{array} \right> = \left( \begin{array}{ccc} \{ j'\} & \{
\Box\!\Box \} & \{ j\} \\ \{ m'\} & \{ \mu \} & \{ m\} \end{array} \right) 
\\
&&\times \sqrt{dim\{ j'\} dim\{ j\} } 
 \left\{ p + \frac{1}{2}
( C_2 (\{ j'\}) - C_2 (\{ j\} ) ) \right\} 
 \left(\begin{array}{ccc} \{ j'\} & \{ \Box\!\Box \} & \{ j\} \\ 
\{ 0\} & \{  0\} &  \{ 0\} \end{array}   \right) , 
\end{eqnarray*}
with all representation labels changed properly as required by the $D \to D+1$
replacement. 

The natural choices for the $D$-vector $X_{a}$ operator is either $A_{a}$
($a=0,1, \dots ,D-1$), 
$$
X_{a} = Q^{(D+1)}_{aD} = J^{(D+1)}_{aD} + T^{(D+1)}_{aD},
$$
or $B_{a}$ ($a=0,1, \dots ,D-1$), 
$$
X_{a} = Q^{(D+1)}_{Da} = J^{(D+1)}_{Da} - T^{(D+1)}_{aD} . 
$$
The above expressions for the $J^{(D+1)}_{aD}$ and $T^{(D+1)}_{aD}$ operators
matrix elements, provide us with an explicit form of the $X^{a}$ operator in
the space of a spinorial field $\Psi (x)$ that transforms w.r.t selected
spinorial $\overline{SL}(D+1,R))$ representation,
\begin{eqnarray*}
&&(i(X_{a})^{B}_{A}\partial_{a} - M) \Psi_{B}(x) = 0 ,\quad A,B = 1/2,
\dots ,\infty ,\\ 
&&\Psi (x) = (\Psi^{(1)}(x), \Psi^{(2)}(x),\dots )^{\rm T} , \quad 
\Psi^{(i)}(x) \to D^{(i)}_{spin}(\overline{SL}(D,R)) \Psi^{(i)}(x) , \\
&&D_{spin}(\overline{SL}(D+1,R)) \supset {\sum_{(i)}}^{\oplus}
D^{(i)}_{spin}(\overline{SL}(D,R)) .
\end{eqnarray*}

Let us finally consider the question of the mass operator $M$. We make use
here of the $s$-parity of the $\overline{SL}(D,R))$ algebra. As stated above,
the $s$-parity of the $Spin(D)$ second-rank tensor representation, $(20\dots
0)$ $=$ $\Box\!\Box$, is $s(\Box\!\Box )= +1$, while the $s$-parity of the
vector representation, $(10\dots 0)$ $=$ $\Box$, is $s(\Box )= -1$. Now, all
the states $\Psi^{(i)}_{A}$ ($A= 1/2,\dots \infty$) of a given spinorial
irreducible representation $D^{(i)}_{spin}(\overline{SL}(D,R))$ are obtained
by consecutive applications of the noncompact $T^{\{ \Box\!\Box \}}$ operators
and therefore have the same $s$-parity. The \\ $s$-parity of the Dirac-like
wave equation $D$-vector, $X^{a} =  X^{\{ \Box \}}$, is $s(X^{a})= -1$, and
thus we find  
$$
<\Psi^{(i)}| X^{a} |\Psi^{(i)}> = 0,\quad i=1,2,\dots \ .
$$ 
We conclude that the mass of an $\overline{SL}(D,R))$-covariant Dirac-like
wave equation can only be of a dynamical origin, i.e. a result of an
interaction. This agrees with the fact that the Casimir operator of the
special affine group $\overline{SA}(4,R)$ vanishes [15] leaving the masses
unconstrained.

\section*{Acknowledgments}

This work was supported in part by MSE, Belgrade, Project-101486.

\end{document}